\documentclass[preprint,12pt]{aastex}
\usepackage{graphicx}
\usepackage{sidecap}
\usepackage{times}
\usepackage{mathptmx}
\usepackage{wrapfig}
\usepackage{titlesec}
\usepackage[breaklinks,colorlinks,urlcolor=gray,citecolor=blue,linkcolor=blue]{hyperref}
\usepackage{ifthen}
\usepackage{afterpage}
\usepackage{appendix}
\usepackage{lineno}
\usepackage{multirow}
\usepackage[dvipsnames]{xcolor, colortbl}
\usepackage{bm}
\usepackage{array}
\newcolumntype{L}[1]{>{\raggedright\let\newline\\\arraybackslash\hspace{0pt}}m{#1}}
\newcolumntype{C}[1]{>{\centering\let\newline\\\arraybackslash\hspace{0pt}}m{#1}}
\newcolumntype{R}[1]{>{\raggedleft\let\newline\\\arraybackslash\hspace{0pt}}m{#1}}

\newcommand{\HI}{H~\textsc{i}}
\newcommand{\HII}{H~\textsc{ii}}

\let\OLDthebibliography\thebibliography
\renewcommand\thebibliography[1]{
  \OLDthebibliography{#1}
  \setlength{\parskip}{0pt}
  \setlength{\itemsep}{0pt plus 0.3ex}
}
\titlespacing{\subsection}{0pt}{2\parskip}{\parskip}
\titlespacing{\subsubsection}{0pt}{1\parskip}{\parskip}
\clearpage
\title{SDSS-V: Pioneering Panoptic Spectroscopy}
\begin{document}

\author{
{\small
%{\bf Management Committee:}
Juna~A.~Kollmeier~(OCIS), Gail~Zasowski~(Utah), Hans-Walter~Rix~(MPIA), Matt~Johns~(UA), \\
Scott~F.~Anderson (UW), Niv~Drory~(UT), Jennifer~A.~Johnson~(OSU), Richard~W.~Pogge~(OSU), Jonathan~C.~Bird~(Vanderbilt), Guillermo~A.~Blanc~(OCIS), Joel~R.~Brownstein~(Utah), Jeffrey~D.~Crane~(OCIS), Nathan~M.~De~Lee~(NKU/Vanderbilt), Mark~A.~Klaene~(APO), Kathryn~Kreckel~(MPIA), Nick~MacDonald~(UCSC), Andrea~Merloni~(MPE), Melissa~K.~Ness~(MPIA), Thomas~O'Brien~(OSU), Jos\'{e}~R.~S\'{a}nchez-Gallego~(UW), Conor~C.~Sayres~(UW), Yue~Shen~(UIUC), Ani~R.~Thakar~(JHU), Andrew~Tkachenko~(KU~Leuven),\\
%{\bf Steering Committee:}
Conny~Aerts~(KU~Leuven),  Michael~R.~Blanton~(NYU), Daniel~J.~Eisenstein~(Harvard), Jon~A.~Holtzman~(NMSU), Dan~Maoz~(TAU), Kirpal~Nandra~(MPE), Constance~Rockosi~(UCSC), David~H.~Weinberg~(OSU), \\
%{\bf Science Working Group Leads:}
Jo~Bovy~(Toronto), Andrew~R.~Casey~(Monash), Julio~Chaname~(Catolica), Nicolas~Clerc~(MPE/CNRS-IRAP), Charlie~Conroy~(Harvard), Michael~Eracleous~(PSU), Boris~T.~G\"ansicke~(Warwick), Saskia~Hekker~(MPS), Keith~Horne~(St~Andrews), Jens~Kauffmann~(MIT--Haystack), Kristen~B.~W.~McQuinn~(UT), Eric~W.~Pellegrini~(Heidelberg), Eva~Schinnerer~(MPIA), Edward~F.~Schlafly~(LBNL), Axel~D.~Schwope (AIP), Mark~Seibert~(OCIS), Johanna~K.~Teske~(OCIS), Jennifer~L.~van~Saders~(OCIS), \\
%{\bf Additional Contributions:}
Andres~Almeida~(U.~de~La~Serena), Brett~H.~Andrews~(Pittsburgh), Maria~Argudo-Fern\'{a}ndez~(Antofagasta), Vladimir~Avila-Reese~(UNAM), Carles~Badenes~(Pittsburgh), John~Bally~(UC--Boulder), Kathleen~Ann~Barger~(TCU), Curtis~Bartosz~(UW), Sarbani~Basu~(Yale), Franz~E.~Bauer~(PUC/MAS/SSI), Rachael~L.~Beaton~(Princeton), Francesco~Belfiore~(UCSC), Eric~C.~Bellm~(UW), Matthew~A.~Bershady~(UW--Madison), Dmitry~Bizyaev~(APO/NMSU), M\'{e}d\'{e}ric~Boquien~(Antofagasta), Mohamed~Bouri~(EPFL), W.~N.~Brandt~(PSU), Jonathan~Brinkmann~(APO), Alyson~Brooks~(Rutgers), Kevin~Bundy~(UCSC), Adam~J.~Burgasser~(UCSD), Joleen~K.~Carlberg~(STScI), Stefane~Caseiro~(EPFL), W.~J.~Chaplin~(Birmingham), Brian~Cherinka~(JHU), Cristina~Chiappini~(AIP), Chia-Hsun~Chuang~(AIP/Stanford), Chris~Collins~(LJMU), D.~Michael~Crenshaw~(GSU), Katia~Cunha~(Steward/ON), Julianne~J.~Dalcanton~(UW), Axel~de~la~Macorra~(UNAM), Aleksandar~M.~Diamond-Stanic~(Bates~College), Francesco~Di~Mille~(LCO/CIS), John~Donor~(TCU), Tom~Dwelly~(MPE), Jason~D.~Eastman~(Harvard/CfA), Eric~Emsellem~(ESO), Xiaohui~Fan~(UA), J.~G.~Fernandez-Trincado~(Concepci\'on/Besan\c{c}on), Diane~Feuillet~(MPIA), N.~Filiz~Ak~(Erciyes), Douglas~P.~Finkbeiner~(Harvard/CfA), Alexis~Finoguenov~(MPE/Helsinki), Peter~M.~Frinchaboy~(TCU), Rafael~A.~Garc\'{i}a~(CEA/AIM), Patrick~Gaulme~(MPS/NMSU), Ortwin~Gerhard~(MPE), Bruce~Gillespie~(JHU), Yilen~G\'{o}mez~Maqueo~Chew~(UNAM), Paul~J.~Green~(SAO), James~E.~Gunn~(Princeton), Daryl~Haggard~(McGill/CIFAR), Patrick~B.~Hall~(York~University), Suzanne~L.~Hawley~(UW), Lynne~Hillenbrand~(Caltech), Natalie~R.~Hinkel~(Vanderbilt), D.~W.~Hoard~(Eureka), David~W.~Hogg~(NYU/MPIA/Flatiron Institute), Eric~Hooper~(UW--Madison), Philipp~H\"{o}rler~(EPFL), Laura~Inno~(MPIA), Amy~M.~Jones~(Alabama), Jean-Paul~Kneib~(EPFL/LAM), Luzius~Kronig~(EPFL), Shri~Kulkarni~(Caltech), David~Law~(STScI), Adam~K.~Leroy~(OSU), Xin~Liu~(UIUC/NCSA), Chelsea~L.~MacLeod~(CfA), Andr\'{e}s~Meza~(U.~Autonoma~de~Chile), Andrea~Miglio~(Birmingham), Ivan~Minchev~(AIP), Dante~Minniti~(UAB/IMA/Vatican~Obs.), John~Mulchaey~(OCIS), Adam~D.~Myers~(Wyoming), Jeffrey~A.~Newman~(Pittsburgh), David~Nidever~(Montana/NOAO), Christian~Nitschelm~(Antofagasta), Stefano~Pasetto~(OCIS), J.~E.~G.~Peek~(STScI), Yingjie~Peng~(KIAA--PKU), Joshua~Pepper~(Lehigh), Marc.~H.~Pinsonneault~(OSU), Adrian~M.~Price-Whelan~(Princeton), M.~Jordan~Raddick~(JHU), Mubdi~Rahman~(JHU), David~Rapetti~(UC--Boulder/NASA--Ames), Florent~Renaud~(Lund~Obs.), Carlos~Rom\'{a}n-Z\'{u}\~{n}iga~(UNAM), Jessie~Runnoe~(Michigan), Jan~Rybizki~(MPIA), David~Schlegel~(LBNL), Donald~P.~Schneider~(PSU), Aldo~Serenelli~(IEEC--CSIC), J.~Michael~Shull~(Colorado), Victor~Silva~Aguirre~(Aarhus), Joshua~D.~Simon~(OCIS), M.~F.~Skrutskie~(UVa), Stephen~A.~Smee~(JHU), Jennifer~Sobeck~(UW), Garrett~Somers~(Vanderbilt), Keivan~G.~Stassun~(Vanderbilt), Matthias~Steinmetz~(AIP), Guy~Stringfellow~(UC--Boulder), Robert~Szabo~(Konkoly~Obs.), Jamie~Tayar~(OSU), Kirill~Tchernyshyov~(JHU), Yuan-Sen~Ting~(IAS/Princeton/Carnegie), Nicholas~Troup~(Salisbury), Juan~David~Trujillo~(UW), Jonathan~Trump~(UConn), Rene~Walterbos~(NMSU), Benjamin~A.~Weaver~(NOAO), Anne-Marie~Weijmans~(St Andrews), Vivienne~Wild~(St~Andrews), Marsha~J.~Wolf~(UW--Madison), Yongquan~Xue~(USTC), Renbin~Yan~(UK), Konstanze~Zwintz~(UIBK) 
}}

\voffset-0.9in
\begin{figure}[t!]
\begin{center}
\includegraphics[width=0.8\textwidth]{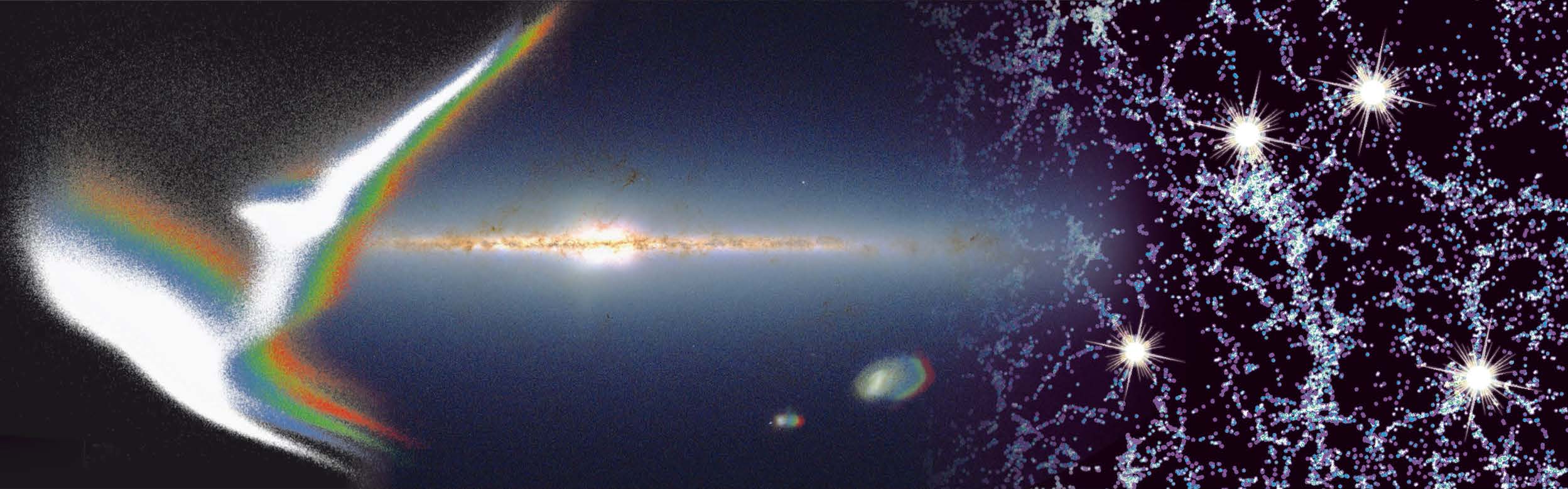}
\end{center}
\end{figure}

%\abstractname{ 
%\begin{abstract}
~
\vspace{-24pt}
\begin{center}
{\bf Abstract}
\end{center}
\vspace{-8pt}
SDSS-V will be an all-sky, multi-epoch spectroscopic survey of over six million objects. It is designed to decode the history of the Milky Way Galaxy (MW), trace the emergence of the chemical elements, reveal the inner workings of stars, and investigate the origin of planets. It will also create an integral-field spectroscopic map of the interstellar gas in the Galaxy and the Local Group that is 1,000 times larger than the current state of the art
and at high enough spatial resolution
to reveal the self-regulation mechanisms of galactic ecosystems. SDSS-V will pioneer systematic, spectroscopic monitoring across the whole sky, revealing changes on timescales from 20 minutes to 20 years.  The survey will thus track the flickers, flares, and radical transformations of the most luminous persistent objects in the universe: massive black holes growing at the centers of galaxies.

The scope and flexibility of SDSS-V will be unique among both extant and anticipated spectroscopic surveys: it is all-sky, with matched survey infrastructures in both hemispheres; it provides near-infrared and optical multi-object fiber spectroscopy that is rapidly reconfigurable to serve high target densities, targets of opportunity, and time-domain monitoring; and it provides optical, ultra-wide-field integral field spectroscopy. 
SDSS-V, with its programs anticipated to start in 2020, will 
be perfectly timed to multiply the scientific output from major space 
missions (e.g., {\it TESS, Gaia}, Spektr-RG--{\it eROSITA}) and ground-based projects. 
SDSS-V builds on the 25-year heritage of SDSS's advances in data analysis, collaboration spirit and infrastructure, and product deliverables in astronomy. The project is now refining its science scope, optimizing the survey strategies, and developing new hardware that builds on the SDSS-IV infrastructure.  We present here an overview of the current state of these developments. SDSS-V is actively seeking to build its consortium of institutional and individual members for a worldwide, partner-driven collaboration.
%\end{abstract}
%}

\section{The Sky is Enormous and It Varies: Why Astrophysics Needs Panoptic Spectroscopy }

The progression of sky surveys that have benefited all areas of astrophysics has, as its logical conclusion, a survey of the entire sky at as many energies as possible.  Since many of the sources in the sky are changing --- in brightness, in color, in position --- multiple epochs of these maps are needed to understand fully the nature of the objects we see.  

Scientifically, there are at least three basic arguments for {\it all-sky} astrophysical surveys. 
\vspace{-8pt}
\begin{itemize} \itemsep -2pt
\item Both the closest sources (e.g., exoplanets and their hosts, low-mass dwarf stars) and the most distant, high-redshift sources are distributed uniformly across the entire sky. Therefore exoplanet research and cosmology, as the cornerstones spanning modern astrophysics, natively rely on all-sky surveys. Whenever one is seeking the best, brightest, or most interesting example of such astrophysical phenomena, it could be anywhere in the sky.
\item The Milky Way Galaxy is the single best ``model organism'' that we have for studying the inner workings of galaxy formation and evolution, as well as an exquisite laboratory in which to study the physics of stars from their birth to their end. The MW is the ultimate all-sky object, where neither Northern nor Southern hemisphere alone can possibly reveal the whole story.
\item The orbital requirements and scanning geometry of multi-year space survey missions naturally enable coverage of the entire celestial globe: from microwaves ({\it COBE}, {\it WMAP}, {\it Planck}), to the infrared ({\it IRAS}, {\it WISE}), optical ({\it Gaia}, soon {\it TESS}), UV ({\it Galex)}, all the way to X-rays ({\it ROSAT}, soon {\it eROSITA}) and gamma rays ({\it Swift, Fermi}). The scientific value of these costly but valuable data sets can be greatly enhanced by ground-based observations, which naturally should then be all-sky.
\end{itemize}
\vspace{-6pt}
To date, all-sky surveys --- from space or from the ground --- have been imaging 
surveys. There has never been a survey providing high-quality spectra across the complete sky\footnote{The {\it Gaia} mission's RVS takes spectra for a subset of bright targets across a limited spectral range.}. {\it The SDSS-V Project now sets out to pioneer this ``panoptic\footnote{panoptic: presenting a comprehensive or encompassing view of the whole} spectroscopy'' by providing the first homogeneous survey of multi-object spectroscopy (MOS) for millions of sources spread across the entire sky.\footnote{http://www.sdss.org/future/}} 

\begin{figure}[!ht]
    \centering
  \includegraphics[width=0.92\linewidth]{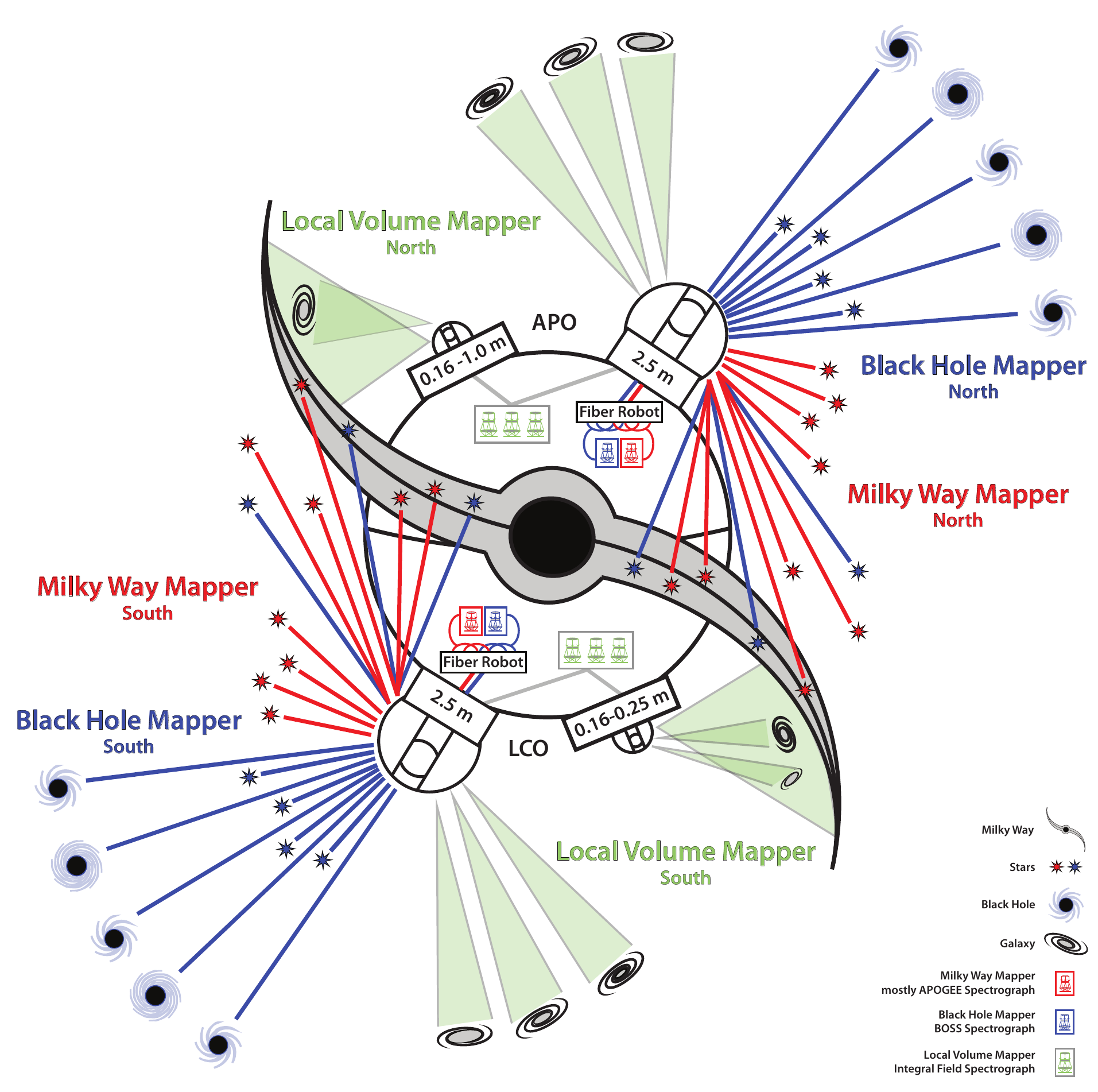}
  \caption{\footnotesize\linespread{1.2} 
  {\bf A schematic representation of SDSS-V:} an all-sky, multi-epoch spectroscopic facility and its science programs. Survey operations will be carried out in two hemispheres, at Apache Point Observatory (APO) and Las Campanas Observatory (LCO). Multi-object fiber spectroscopy will be obtained with two 2.5m telescopes, each feeding a near-IR APOGEE spectrograph (300 fibers, $R \sim 22,000$) and an optical BOSS spectrograph (500 fibers, $R \sim 2,000$); this configuration enables a sky-survey rate of $\sim$40~deg$^2$~hr$^{-1}$. Ultra-wide field integral field spectroscopy will be performed mostly with smaller telescopes at these observatories, with $\sim$2,000-fiber bundles feeding three optical spectrographs in each hemisphere. This schematic also outlines the three primary science programs: the Milky Way Mapper, drawing on both APOGEE (red) and BOSS (blue) spectra; the Black Hole Mapper, acquiring BOSS spectra of fainter targets; and the Local Volume Mapper, performing IFU mapping of the ionized ISM in the MW and nearby galaxies.
}
  \label{fig:SDSSV_schematic}
\end{figure}

Furthermore, many sources in the sky (perhaps most, if one is looking closely enough) are changing with time, moving through space and/or changing flux and color for a wide range of interesting physical reasons. Examples of periodic phenomena include planets that occult their hosts, stars that oscillate, and close binary components that deform each other. Transient phenomena encompass the incessant ringing and death throes of stars, magnetic flaring, and variations in black hole accretion rates. 
The time domain has been recognized as one of the great astrophysical frontiers of our time; this interest is reflected in the enormous investments in the space missions {\it Kepler}, {\it Gaia}, {\it TESS}, and {\it PLATO}, and the ground-based surveys PTF, PS1, ZTF, and LSST. At heart, all of these projects are time-domain imaging surveys. There has not been a comparably coherent effort to systematically explore and exploit the time domain with spectroscopy. {\it The second key aspect of SDSS-V involves expanding its panoptic spectroscopy to include multi-epoch survey spectroscopy.}

Further, many astrophysical phenomena do not lend themselves to being parsed into sets of discrete sources to be observed by multi-object spectroscopy. These phenomena call for contiguous spectral mapping, or integral-field spectroscopy (IFS), over a wide range of angular and physical scales. Pushing IFS to the all-sky regime is even more daunting than all-sky MOS, as evidenced by the fact that the largest contiguous (optical) IFS maps of the sky cover only 0.001\% of it.
{\it SDSS-V's third central goal is to bring IFS to a regime of mapping an appreciable portion ($>$3000~deg$^2$, $\sim$8\%) of the sky with optical data-cubes.}

SDSS-V will enable high-quality, near-IR and optical MOS observations of over $6\times 10^6$ objects across the entire sky, with homogeneous multi-epoch spectroscopy for about one million of them, and it will carry out ultra-wide-field IFS mapping across more than 3000~deg$^2$ of the sky. These advances can only be achieved with a suitable combination of hardware and survey strategy. They require wide-field telescopes and the possibility to re-acquire new spectroscopic targets rapidly (e.g., every 15~minutes), with a focus on bright sources for MOS and on emission lines for IFS. SDSS-V will meet these requirements by building on the infrastructure, instrumentation, survey operational expertise, and collaboration heritage of the Sloan Digital Sky Surveys I-IV \citep[SDSS;][]{York_2000_sdss1,Eisenstein_etal_2011,Blanton_etal_2017}, as well as constructing new hardware based on proven technology (Figure~\ref{fig:SDSSV_schematic}).

This extensive survey infrastructure will enable the first panoptic spectroscopic survey, comprising three primary survey programs (Section~\ref{sect:TheThreeSurveys}). The {\it Milky Way Mapper} (MWM; Section~\ref{sec:mwm}) will provide a) a global spectroscopic map of the MW, using near-IR MOS concentrated at low Galactic latitudes; b) a multi-epoch stellar astrophysics survey, focused on interesting targets of {\it Gaia} and TESS; and c) a multi-epoch survey of young, massive stars throughout the Galaxy. The {\it Black Hole Mapper} (BHM; Section~\ref{sec:bhm}) will focus on long-term, time-domain studies of AGN, including direct measurement of black hole masses and changing-look quasars, and on the optical characterization of {\it eROSITA} X-ray sources. The {\it Local Volume Mapper} (LVM; Section~\ref{sec:lvm}) will provide the first integral field spectral map, spanning the full optical window, of the bulk of the MW disk, the Magellanic Clouds, the Andromeda Galaxy, and other galaxies in the Local Volume.  Table~\ref{tab:SDSSV_program_table} presents the basic survey parameters of the three {\it Mappers}.

\begin{table}[h]
\large
\scalebox{0.75}{
\begin{tabular}{|C{3.0cm}||L{3.5cm}|m{3.5cm}|L{5cm}|m{5cm}|}
 \hline
 \rowcolor{SkyBlue}
{\bf Program} & {\bf Science Targets} & {\bf N$_{\rm Objects}$ and/or \newline Sky Area} & {\bf Primary Spectral Range and Hardware} & {\bf Primary Science Goals} \\
\hline
\hline
{\bf M}ilky {\bf W}ay {\bf M}apper (MWM) & Stars across the Milky Way & $>$6M stars; all-sky & IR; APOGEE ($R\sim 22,000$) with fiber-positioning system & Understanding the formation of the Milky Way and the physics of its stars \\
\hline
{\bf B}lack {\bf H}ole {\bf M}apper (BHM) & Primarily supermassive black holes & $>$400,000 sources; all-sky & Optical; e.g., BOSS ($R\sim 2000$) with fiber-positioning system & Probing black hole growth and mapping the X-ray sky \\
\hline
{\bf L}ocal {\bf V}olume {\bf M}apper (LVM) & ISM \& stellar populations in the MW, Local Group, and nearby galaxies & $>$25M contiguous spectra over 3,000 deg$^2$ & Optical; new integral field spectrographs covering 3600-10000\AA\, at $R\sim 4000$ & Exploring galaxy formation and regulation by star formation; feedback, enrichment, \& ISM physics \\
\hline
\end{tabular}
}
\caption{\footnotesize\linespread{1.2}{
A summary of the SDSS-V Mapper programs: Milky Way Mapper, Black Hole Mapper, and Local Volume Mapper.
}}
\label{tab:SDSSV_program_table}
\end{table}

In Section~\ref{sect:SurveyImplementation} we describe the survey implementation, while the upcoming preparatory developments and the collaboration building is discussed in Section~\ref{sect:DevelopmentCollaboration}. We provide an outlook of the survey prospects in Section~\ref{sec:future}.

\section{The SDSS-V Mapper Components}\label{sect:TheThreeSurveys}

\subsection{Milky Way Mapper}
\label{sec:mwm}

The ecosystem of stars, gas, dust, and dark matter in large galaxies like the MW has been shaped over billions of years by a variety of physical processes that operate across an enormous range of physical scales. Despite this complexity in galaxy formation, we now observe a population of galaxies that is very ordered, despite galaxy masses spanning from hundreds of stars to a hundred billion stars. 
Explaining how such regularity emerges in a cosmological context from such complex and varied physics is a central challenge of modern astrophysics.  The Milky Way Mapper (MWM) survey will exploit our unique perspective within our Galaxy to address this issue by creating a unique global Galactic map that encompasses the evolutionary record contained in its stars and interstellar material (ISM). 

\begin{figure}[htbp]
    \centering
  \includegraphics[trim=0in 3.25in 0in 3.75in, clip, width=\textwidth]{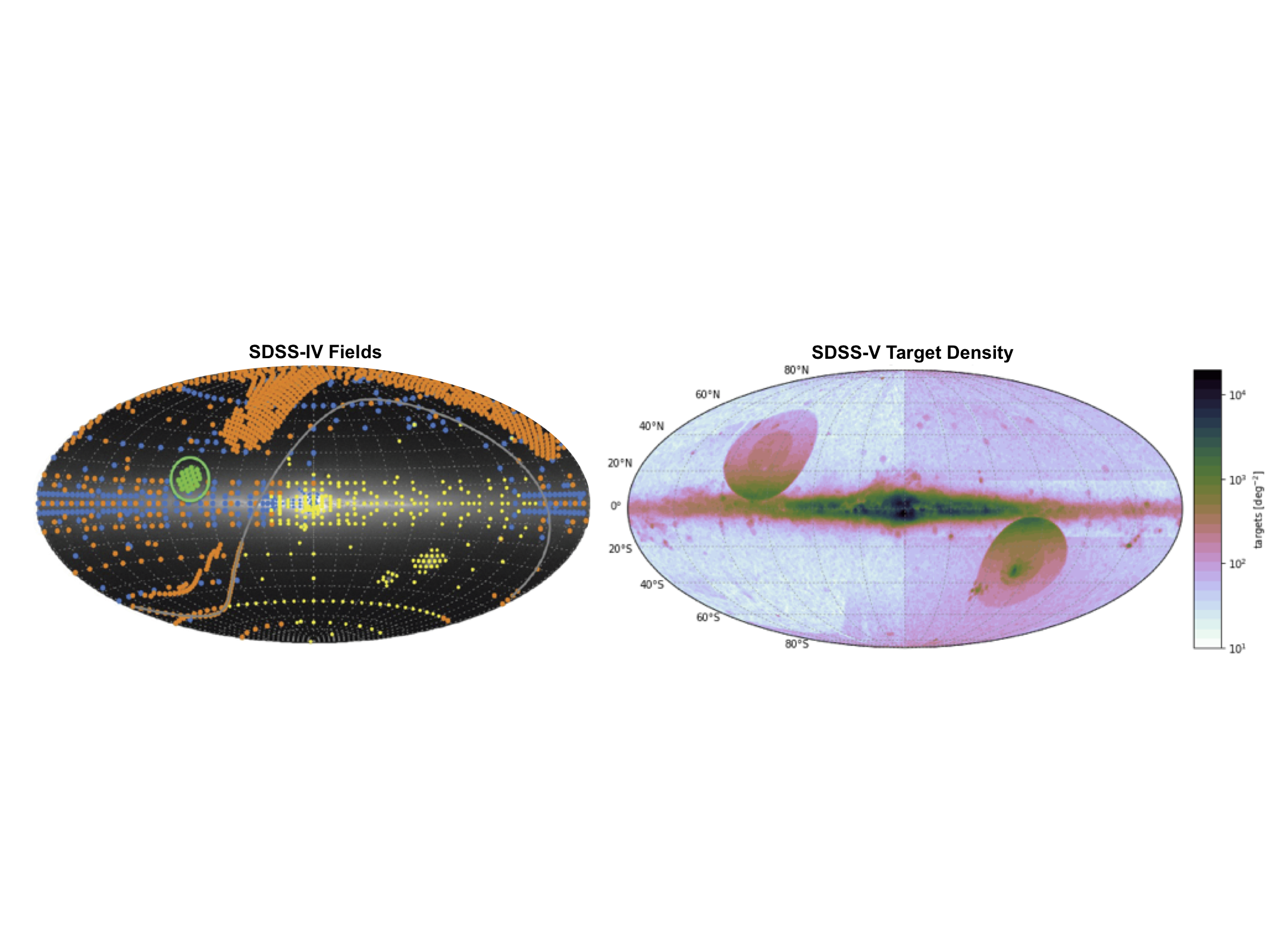}
  \caption{\footnotesize\linespread{1.2} 
  {\bf Evolution of SDSS on-sky target density:} Left: SDSS-IV field map, in Galactic coordinates. The colored dots show regions of the sky targeted by SDSS-IV's MaNGA, eBOSS, and APOGEE-2 (orange) and APOGEE-2 only (yellow, green, and blue).  There are no data where the background image is visible. Right: Density map of SDSS-V's spectroscopic targets (objects per square degree). The analysis of the data from the sparse but deep sampling provided by earlier generations of SDSS allows us to exploit new technologies and analysis techniques to cover the entire sky contiguously with spectra in SDSS-V.  
}
  \label{fig:SDSSV_programs}
\end{figure}

To truly utilize our MW as a galaxy model organism, surveys must trace the entire hierarchy of structure within the Galaxy with maps that are global, contiguous, and densely sampled throughout the (largely dust-obscured) disk- and bulge-dominated regions of the stellar MW. High quality spectra of stars are rich with information about their basic physical parameters, including their ages, chemistry, and kinematics. Interstellar spectral lines probe the composition and dynamics of the MW's gas and dust, from which new stars are forming. These properties provide the best approaches to quantitatively test models of the most uncertain galaxy formation physics \citep[][]{RixBovy2013,Gerhard2016}. 
Within the MWM program, the Galactic Genesis Survey (GGS) will produce the first spectroscopic stellar map that is contiguous, densely sampled, and all-sky but focused on the low Galactic latitudes where most stars lie, and which includes detailed information on {\it each} star and its foreground ISM. With its near-IR, multi-epoch spectroscopy through the entire Galactic plane, SDSS-V will also significantly expand the spectroscopic census of young stars in the MW, characterizing their masses, ages, multiplicity, etc., thus painting a global picture of the ``recent Galaxy.''

\begin{figure}[ht]
    \centering
  \includegraphics[width=1.0\linewidth]{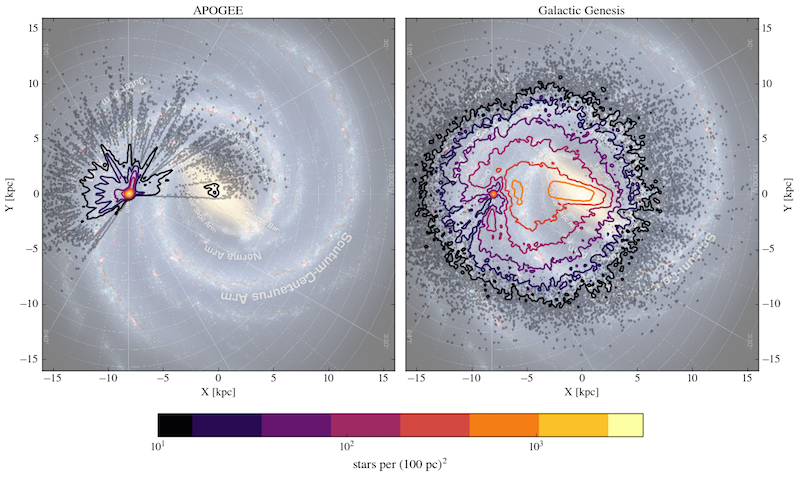}
  \caption{\footnotesize\linespread{1.2}
  {\bf Evolution of SDSS in-plane Galactic target density:} Midplane target surface density of the recent APOGEE DR14 catalog (left) and MWM's Galactic Genesis Survey (GGS; right). The maps show a face-on schematic of the Milky Way ({\it credit: NASA/JPL-Caltech/R. Hurt}) beneath target density contours. The Sun is located 8~kpc from the center of the Galaxy, at ($X, Y = -8.0, 0.0$). Light gray contours show areas with observed/anticipated stars at surface densities $<$10 per (100 pc)$^2$; colored contours follow the colorbar. These contours only contain stars within 500~pc of the midplane, summing to $1.5 \times 10^5$ in APOGEE DR14 and $3.6 \times 10^6$ stars in GGS. For APOGEE, we show stars with distances reported in the APOGEE DR14 Distance Value Added Catalog, which represent $\sim$95\% of all main survey targets. We note that ongoing APOGEE-2 observations will fill in the fourth quadrant of the Galaxy. Distance distributions for SDSS-V targets were calculated using a mock GGS observation of the Galaxia model of the MW \citep{Sharma2011} and a 3D extinction map \citep{Bovy2016}. 
  }
  \label{fig:midplane_GG}
\end{figure}

MWM will take advantage of several factors to produce this remarkable data set: {\it Gaia} photometry and astrometry, the all-sky coverage of SDSS-V, the rapid target allocation enabled by the robotic fiber positioner (Section~\ref{sec:infrastruct_instrument}), the APOGEE spectrographs' IR wavelength coverage and resolution, the large FOV of the APO and LCO telescopes (Section~\ref{sec:infrastruct_instrument}), and novel spectral analysis techniques. GGS's rapid, wide-angle survey mode is enabled by its focus on bright ($H<11$), yet intrinsically luminous (and thus distant) sources. These include variable star distance indicators such as Cepheids and Mira variables \citep[cataloged by, e.g., VVV;][]{Minniti_2010_VVV}, which fall within GGS's magnitude limits even when in the disk beyond the bulge.  GGS's immediate product will be a Galactic census of stellar orbits, ages, and detailed abundances as a function of three-dimensional position across the entire Milky Way disk and bulge. GGS will collect spectra from more than 5 million stars across the full sky, most of them from a contiguous area of $\gtrsim$3,000 deg$^2$ in the Galactic midplane (Figures~\ref{fig:SDSSV_programs} and \ref{fig:midplane_GG}; Table~\ref{tab:ggsa_classes}).  These data will provide the means to address numerous long-standing questions, including the dominant formation mechanisms of the MW, hierarchical accretion, radial migration, and the place of the MW in a cosmological context.

\begin{figure}[!hptb]
    \centering
  \includegraphics[width=\linewidth]{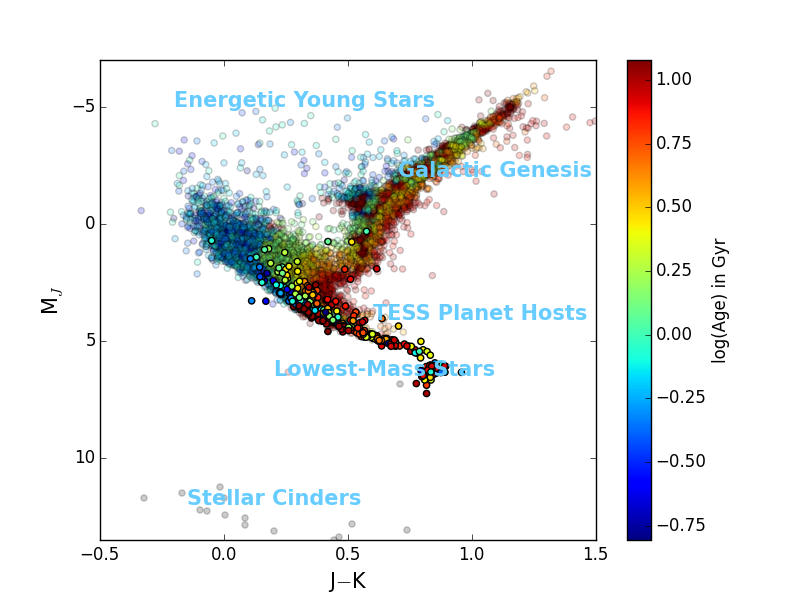}
  \vspace{-1cm}
  \caption{\footnotesize\linespread{1.2}
  {\bf Stellar astrophysical targets in the MWM:} The $(J-K)$ color and absolute $J$ magnitude of 0.001\% of the $\sim$1 billion stars that {\it Gaia} will observe, color-coded by their expected ages based on a Besan\c{c}on Galaxy model \citep{robinetal2003}. The wide range of ages of the red giants provides a perfectly-suited exploration space for the MWM (Section~\ref{sec:mwm}), given that we can determine their asteroseismic-calibrated ages. The luminous hot stars in the upper left ionize the gas seen by the LVM in the Milky Way (Section~\ref{sec:lvm}), and the cool dwarfs on the lower right yield prime hunting ground for rocky planets in the habitable zone, whose host stars must be carefully characterized. The stars marked in bright colors represent those that are within 100 pc of the Sun and therefore part of the MWM's solar neighborhood census. The lowest-mass stars will be a major component of this census, especially since subsequent {\it Gaia} catalogs will have distances for much fainter stars than Data Release 1 does. The gray points with $M_J>10$ mark white dwarfs with {\it Gaia} DR1 distances; the number of these ``cinders'' of low-mass stars will increase by a factor of 10$^5$ as {\it Gaia} continues. With knowledge of the white dwarf initial mass-final mass relation, ages can determined for these as well.}
  \label{fig:hr}
\end{figure}

The rigorous interpretation of GGS's data will ultimately rely on understanding the complete lifecycle of stars from birth to death, including multi-star systems \citep[e.g.,][]{Raghavan2010,DK2013}.  However, many questions remain about these lifecycles: What determines the mass and multiplicity of stellar systems?  How does multiplicity affect stellar evolution? What is the relationship between stellar and planetary properties \citep[e.g.,][]{brugamyeretal2011,Adibekyan2013}?  How does nucleosynthesis proceed throughout the lifetimes and death throes of different kinds of stars?

\begin{figure}[!hptb]
    \centering
  \includegraphics[trim=0.1in 0.4in 0.3in 0.6in, clip, width=0.9\textwidth]{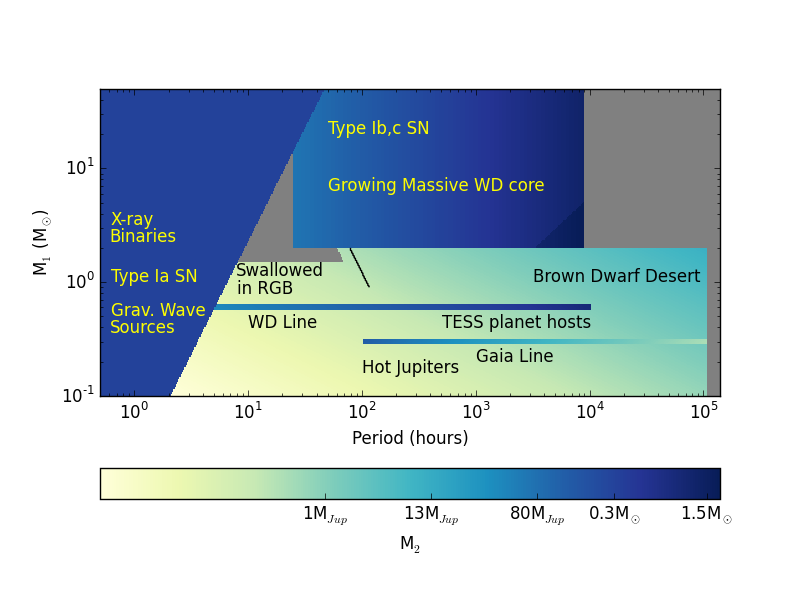}
  \caption{\footnotesize\linespread{1.2}
  {\bf SDSS-V's stellar companion mass sensitivity:} The minimum companion mass SDSS-V can detect in systems with a range of periods and primary masses, in the context of several scientifically interesting regimes. For example, the left-most region has a high minimum mass because systems there have undergone common envelope evolution, and there must be at least one white dwarf to survive. 
We have indicated the area where stars becoming red giants swallow their closest companions. Faint white dwarfs will need to have $\sim$15~km~s$^{-1}$ RV variability to be detected by the optical BOSS spectrographs. This precision is well matched to the RV amplitudes of several 100~km~s$^{-1}$ in double-white dwarf binaries. The ``WD Line'' shows the minimum mass of an unseen companion around a WD, and we can clearly detect neutron star and black hole companions out to a period of many months. The ``{\it Gaia} Line'' shows the minimum secondary mass detectable by {\it Gaia} around a 0.2 M$_{\odot}$ star at a distance of 250 pc. This illustrates how {\it Gaia} and SDSS-V complement each other:
astrometry becomes increasingly powerful at long periods, spectra at short periods.
Gray areas will be explored statistically, but we will not have full orbital information for systems in these regions.}
  \label{fig:binary}
\end{figure}

The key data to understanding such outstanding questions of stellar astrophysics are long duration, high precision, time-series photometry and spectroscopy of large, carefully-selected samples of single- and multiple-star systems.  In particular, asteroseismology and {\it absolute} flux measurements \citep[e.g., from {\it Kepler} and {\it TESS}, and {\it Gaia}, respectively; see][]{Campante_2016_TESSseismology,Huber_2017_TGASseismology,Stevens_2017_GaiaAbsFlux} have recently emerged as game-changers in terms of deepening our understanding of stellar astrophysics---literally, by letting us peer past the previous limit of stellar photospheres.  SDSS-V will use its panoptic spectral capabilities for a comprehensive investigation of Stellar Astrophysics (SA) and of Stellar System Architecture (SSA) over a range of 10$^4$ in the masses of stars that belong to binaries, 0.5 hours to $>$12 years in orbital period, and a few pc to $>$15 kpc in distance from the Sun.  The overarching goal of these programs is to consistently and comprehensively measure mass, age, chemical composition, internal structure, rotation, and the presence of companions for vast samples of stars across the color-magnitude diagram (Figure~\ref{fig:hr}).

The SA program will include observations necessary for precise age measurements of giant stars with asteroseismic detections \citep[e.g.,][]{Martig2016}; observations of massive stars to constrain the true relationships between masses, radii, rotation, and internal mixing; observations of several thousands of white dwarfs to improve our understanding of their evolution and mass return to the ISM; observations of deeply embedded stellar clusters, caught in the act of forming stars at numerous stages; and observations of a volume-limited sample of stars within $\sim$100~pc.  The SSA program will target tens of thousands of multiple-body systems across a diverse range of Galactic environments, with a wide range of periods and primary masses (Figure~\ref{fig:binary}).  SSA seeks to explain the dependence of multiplicity on stellar mass and environment, the frequency and properties of binary systems with compact objects that give rise to explosive events and gravitational waves \citep[e.g.,][]{LIGO_2017_GW170817}, and the effect of host system composition on exoplanet frequency and habitability.  See Table~\ref{tab:ggsa_classes} for more details on targeting for the SA and SSA programs.

\begin{table}[!h]
\scalebox{0.8}{
\begin{tabular}{ |p{1.9cm}|p{5.3cm}|p{1.5cm}|p{1.2cm}|p{8.0cm}|  }
 \hline
 \multicolumn{5}{|c|}{{\bf Galactic Genesis \& Stellar Astrophysics Targeting Classes}} \\
 \hline
{\bf Instrument} & {\bf Selection} & {\bf N$_{\rm Targets}$} & {\bf N$_{\rm Epochs}$} &  {\bf Comments} \\
 \hline
 \multicolumn{5}{|l|}{\textcolor{violet}{Galactic Genesis Survey:} mapping the dusty disk}\\
 \hline
APOGEE& $H$$<$11, $G-H>3.5$ & 4,800,000 & 1 & dust-extinguished disk \\
   APOGEE & $|z| < 200$ pc, $H$$<$11,  d$<$5~kpc & 125,000 & 1 & to complete high-res ISM map \\
 \hline
\multicolumn{5}{|l|}{\textcolor{violet}{Binaries with Compact Objects:} enumerating the populations of binaries with white dwarfs, neutron stars, or black holes,} \\
\multicolumn{5}{|l|}{selected by variability} \\
\hline
 BOSS & PTF, ZTF, {\it Gaia} variability & 30,000 & 3 & binaries with WDs, NSs, and BHs   \\
 BOSS  & {\it Gaia} parallaxes& 30,000  & 1 & wide WD+MS/RGB binaries    \\
  \hline
\multicolumn{5}{|l|}{\textcolor{violet}{Solar Neighborhood Census:} observing all stars within 100 pc, giving the best probe of low-mass stars, whether in single or} \\
\multicolumn{5}{|l|}{binary systems}\\ 
\hline
 APOGEE, BOSS & \multirow{2}{*}{d$<$100 pc, $G<20$, $H<12$} & \multirow{2}{*}{400,000} & \multirow{2}{*}{2} & \multirow{2}{*}{1000$\times$ increase in volume \& stars} \\
  \hline
\multicolumn{5}{|l|}{\textcolor{violet}{White Dwarf Chronicle:} using white dwarfs and their evolved companions to measure the SFH and age-metallicity relation }  \\
\hline
 BOSS & $G<20$ & 300,000 & 3 & 15$\times$ increase in sample size\\
  \hline
\multicolumn{5}{|l|}{ \textcolor{violet}{TESS Exoplanet Host Candidates:} observing all TESS short-cadence targets in the CVZs } \\
\hline
 APOGEE  & $H \leq 13.3$ & 300,000 & 1--8 & all short-cadence targets \& planet hosts\\
 \hline
\multicolumn{5}{|l|}{ \textcolor{violet}{Binaries Across the Galaxy:} measuring environmental dependence of binary fraction in the disk, bulge, halo, and stellar} \\
\multicolumn{5}{|l|}{clusters; probing the brown-dwarf desert beyond solar-type stars }\\
\hline
 \multirow{2}{*}{APOGEE} & $H$$<$13.4, N$_{\rm Epoch} \geq 6$ by the start of SDSS-V  & \multirow{2}{*}{60,000} & \multirow{2}{*}{6--18} & gives orbits with 24--40 epochs for all targets with long APOGEE baselines \\
 \hline 
\multicolumn{5}{|l|}{\textcolor{violet}{{\it Gaia} Astrometric Binaries:} characterizing rare systems that have good astrometric orbits but limited other information, } \\
\multicolumn{5}{|l|}{from {\it Gaia}'s sample of $>10$ million stars} \\
\hline
  APOGEE, BOSS & \multirow{2}{*}{$d<3$ kpc} & \multirow{2}{*}{200,000}  & \multirow{2}{*}{1} & \multirow{2}{*}{rare types of systems} \\
 \hline 
\multicolumn{5}{|l|}{\textcolor{violet}{TESS Red Giant Variability:} measuring spectroscopic properties for red giants in TESS that have seismic and/or granulation} \\
\multicolumn{5}{|l||}{lightcurve signatures} \\
\hline
 APOGEE & $H<12.5$ & 250,000 &  1 & stars with at least 80 days of TESS observation \\
 \hline
\multicolumn{5}{|l|}{\textcolor{violet}{Massive, Convective Core Stars:} combining dynamic and asteroseismic measurements of binary OBAF stars in the TESS CVZs} \\
\multicolumn{5}{|l|}{and characterizing their multiplicity} \\ 
\hline
APOGEE & \multirow{2}{*}{$H<12$} & 200,000 & 2 & detection of single vs. binary systems\\
   APOGEE & & 500 & 25 & $>$10$\times$ increase in current sample size \\
 \hline
\multicolumn{5}{|l|}{\textcolor{violet}{Young Stellar Objects:} quantifying the stellar populations in star-forming regions, including identifying sources of ionizing} \\
\multicolumn{5}{|l|}{radiation and characterizing the binary frequency} \\
\hline
APOGEE & $H<12$, $d<1$ kpc & 20,000 & 12 & nearby star-formation regions \\
    APOGEE & $H<12$ & 3,500 & 8 & high-mass star-formation regions\\
  APOGEE & $H<12$, $|b|<2^\circ$ & 10,000 & 2 & massive young stars in the Galactic Plane \\
   APOGEE & $H<13$ & 10,000 & 2 & Central Molecular Zone \\
 \hline 
\end{tabular}
}
\caption{\footnotesize\linespread{1.2}{Current targeting strategy for the subsamples in the GG, SA, and SSA components of the MWM (Section~\ref{sec:mwm}). Note that the targeting details are still being optimized.}}
\label{tab:ggsa_classes}
\end{table}

\subsection{Black Hole Mapper}
\label{sec:bhm}

Quasars/AGN\footnote{Historically, ``quasars'' and ``Active Galactic Nuclei (AGN)'' describe different classes of objects, but here we use them interchangeably in recognition of the fact that they both describe accreting supermassive black holes \citep[e.g.,][]{Merloni_2016}.} are among the most luminous objects in the Universe. Powered by accretion onto supermassive black holes (SMBHs) at the centers of galaxies, quasars are beacons marking and tracing the growth of black holes across cosmic distance and time. The tight correlations between the masses of these central SMBHs and the properties of their hosts \citep[e.g.,][]{Kormendy_Ho_2013} demonstrate a clear connection between the formation of the stellar component of a galaxy and the growth of its central BH. This connection means that quasar studies are not only critical for understanding SMBHs and their accretion physics, but are also closely linked to galaxy formation and evolution being explored by SDSS-V's MWM (Section~\ref{sec:mwm}) and LVM (Section~\ref{sec:lvm}). 

\begin{figure}[hb!]
%\vspace{0.2in}
\centering
  \includegraphics[width=1.0\linewidth]{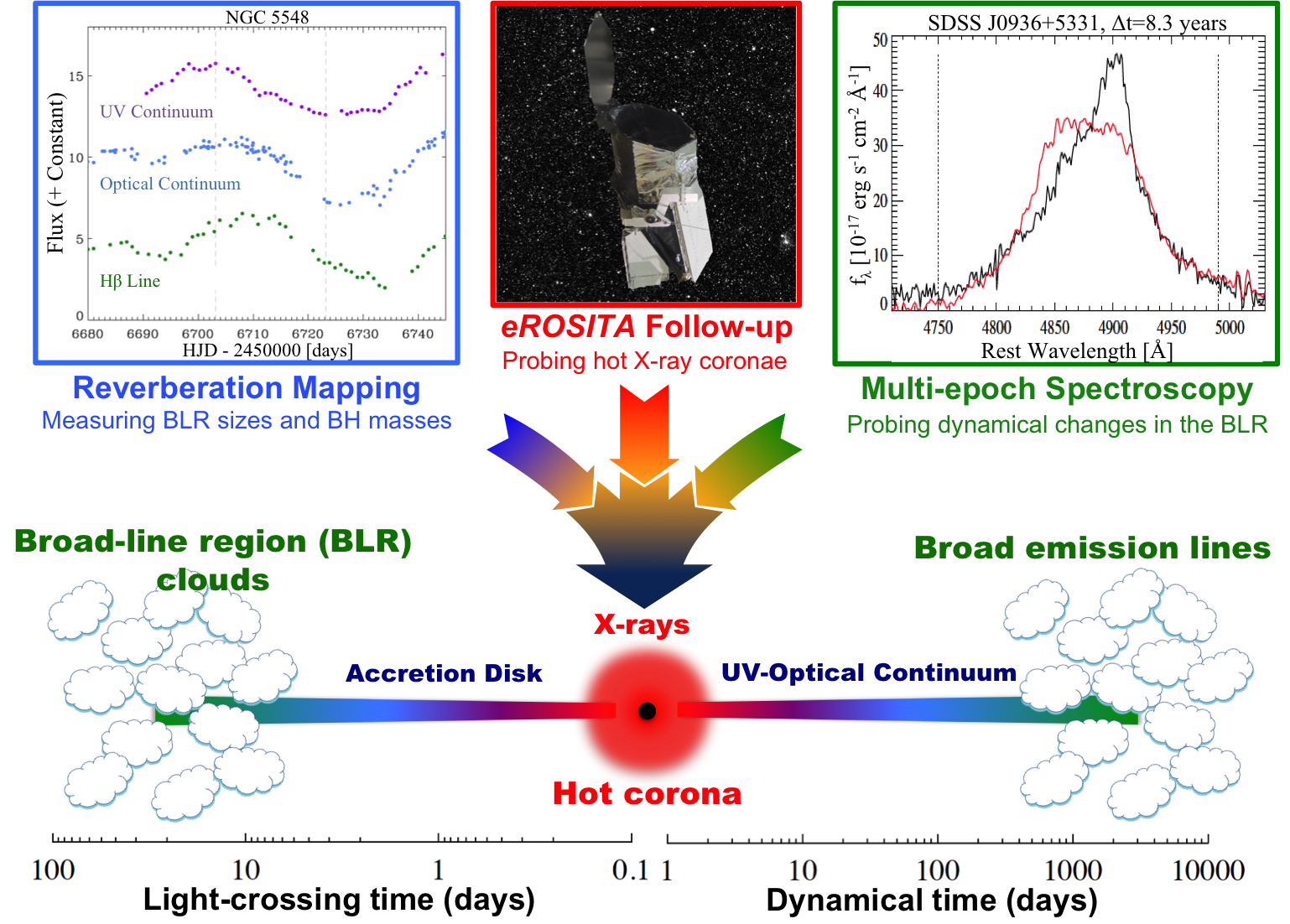}
  \caption{\footnotesize\linespread{1.2}
  {\bf Schematic of the innermost regions around a quasar's
central supermassive black hole (BH):} the X-ray corona, accretion disk, and broad-line region (BLR). SDSS-V will explore the physics of supermassive BH accretion and dynamics with three parallel approaches: reverberation mapping, {\it eROSITA} follow-up, and multi-epoch spectroscopy (top three panels). Top left: an example of time delays \citep{Pei_etal_2017} between UV/optical continua from the accretion disk and emission line flux from the BLR, which yields the BH mass. Top center: the {\it eROSITA} X-ray telescope that, combined with SDSS-V spectra, will conduct a census of X-ray/optical properties for $>10^5$ quasars.  Top right: an example of multi-epoch spectra \citep{Liu_etal_2014} that reveal a marked change in the broad-line profile for a quasar over a rest-frame time baseline of 8.3 years.  This change, similar to those the BHM will provide for large numbers of quasars, is a probe of dynamical processes within the BLR.}
\label{fig:agn_schematic}
\end{figure}

Quasar variability, which can occur across the entire electromagnetic spectrum and on time scales of hours to decades, encodes information on the structure, dynamics, and evolution of emitting regions, even in areas too small to resolve spatially with any telescope for the foreseeable future.  Nearly all quasars also produce energetic X-ray emission, largely from these inner regions close to the SMBH, that is generally less affected by intervening obscuration and thus provides diagnostics complementary to those at longer wavelengths. Observational tests of BH/quasar theory therefore require three primary measurements: precise mass constraints, multi-wavelength SEDs, and a detailed characterization of variability (Figure~\ref{fig:agn_schematic}). SDSS-V's Black Hole Mapper (BHM) will provide these measurements for a large sample of quasars by adding wide-area, multi-epoch optical spectroscopy (Table~\ref{tab:bhm_target_classes}) to the current and upcoming time-domain optical imaging projects and to the next-generation of X-ray surveys (e.g., ZTF, LSST, and {\it eROSITA}). 

\begin{table}[h]
\scalebox{0.9}{
\begin{tabular}{ |m{6.0cm}|m{3.8cm}|m{3.0cm}|m{1.4cm}|m{1.4cm}| }
 \hline
 \multicolumn{5}{|c|}{{\bf SDSS-V Black Hole Mapper Targeting}} \\
 \hline
  {\bf Science Goals}  & {\bf Primary Selection} & {\bf Density [deg$^{-2}$]} & {\bf N$_{\rm targets}$} & {\bf N$_{\rm epochs}$}\\
 \hline 
 \hline
 Reverberation mapping, \newline BH masses & Optical QSOs, $i<20$ & 30--50 & 1,500 & 174 \\
 \hline
BH accretion and outflow astrophysics, changing look quasars  & Optical QSOs, $i<19$ & 10 & 25,000 & 3--13 \\
 \hline
{\it eROSITA} follow-up, AGN, X-ray binaries, galaxy clusters  & $f_{\rm X-ray} \geq 2.5\times 10^{-14}$ erg~s$^{-1}$~cm$^{-2}$, $i<21.5$ & 20--50 & 400,000 & 1--3 \\
 \hline
\end{tabular}
}
\caption{\footnotesize\linespread{1.2} 
Scope of the BHM program to be carried out with, e.g., optical BOSS spectrographs in two hemispheres. 
}
\label{tab:bhm_target_classes}
\end{table}

The past few decades have witnessed the success of using quasar time-domain variability to constrain basic models of quasars. A prime example is reverberation mapping\footnote{In RM, we measure the time delay between variability in the ionizing continuum from the accretion disk and its ``echo'' from the BLR \citep[e.g.,][]{Blandford_McKee_1982,Peterson_1993}.} (RM) of the broad emission line region (BLR). RM delays measure the typical sizes of the BLR and, when combined with the velocity width of the broad emission lines, allow a virial estimate of the BH mass, the most fundamental of all BH parameters. Large, representative samples of quasars with robust spectroscopic variability studies have not yet been assembled \citep[e.g.,][]{Vestergaard_2011}, and the power of {\it spectral} variability to constrain quasar models has been insufficiently explored.

\begin{figure}[!ht]
\centering
 \includegraphics[trim=0cm 1cm 0cm 0cm, clip, width=0.95\linewidth]{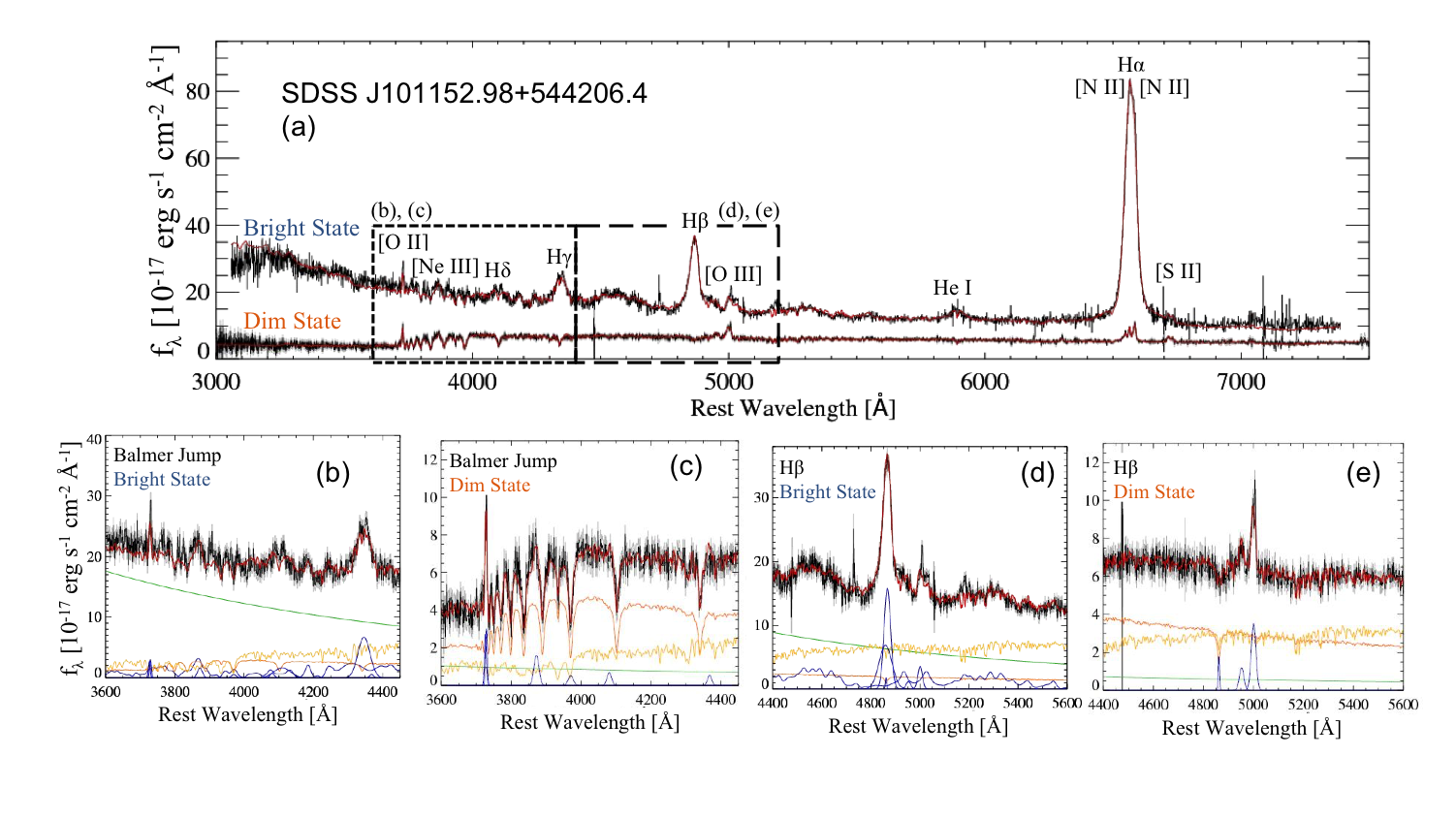}
 \vspace{-0.2in}
  \caption{\footnotesize
{\bf Spectral comparison of two epochs for a ``changing look'' quasar
(CLQ):}
(a) SDSS~J101152.98+544206.4, identified in the TDSS \citep[][]{Runnoe_etal_2016}, is
a $z=0.246$ quasar that dimmed by $\sim$0.5~mag over about 500 days in the rest frame (MJDs 52652--57073), especially in the blue part of the spectrum.  In the lower panels, spectral decomposition is shown for the bright and dim epochs around the Balmer jump region (panels (b) and (c), respectively) and for the H$\beta$/[O\,III]/Fe\,II region (panels (d) and (e)).  
Colors denote data (black), uncertainties (gray), total best-fitting model (red), old and young stellar templates (yellow, orange), power-law continuum (green), and emission lines (blue).  In the dim state, the host stellar populations are highly prominent, while the broad emission lines have
nearly disappeared. SDSS-V quasar time-domain spectroscopy will systematically probe for such surprising accretion-state transitions in a far larger survey of quasars.
}
\label{fig:clagn}
\end{figure}

SDSS-III and -IV demonstrated the potential of this approach \citep[e.g.,][]{Shen_etal_2015a,Grier_2017_RMsdss4}, and SDSS-V's BHM will exploit this potential on ``industrial scales.'' Spectroscopic RM sampling hundreds of epochs to determine precise BH masses will be performed for $\sim$1,000--1,500 quasars with a range of redshifts ($0.1<z<4.5$) and luminosities ($L_{\rm bol}\sim 10^{45}-10^{47}\,{\rm erg\,s^{-1}}$), an increase of $\sim$25$\times$ the historical sample of nearby, low-luminosity AGN with RM-determined masses. With a more modest number of epochs (a few to a dozen per target, yielding final baselines spanning months to a decade), BHM will also characterize the optical spectral variability of approximately 25,000 quasars, illuminating the astrophysics of SMBH accretion disks, dynamical changes in the BLRs, signatures of binary BHs, and the properties of quasar outflows. In addition, studies of ``changing-look'' AGN (in which the broad lines around the AGN either appear or disappear; Figure~\ref{fig:clagn}) comprise a burgeoning field that challenges standard accretion disk theory \citep[e.g.,][]{LaMassa_2017_changinglookAGN}; many of these intriguing sources will also be discovered in the BHM's repeat spectroscopy program.

 \begin{figure}[!ht]
 \centering
  \includegraphics[width=0.9\linewidth]{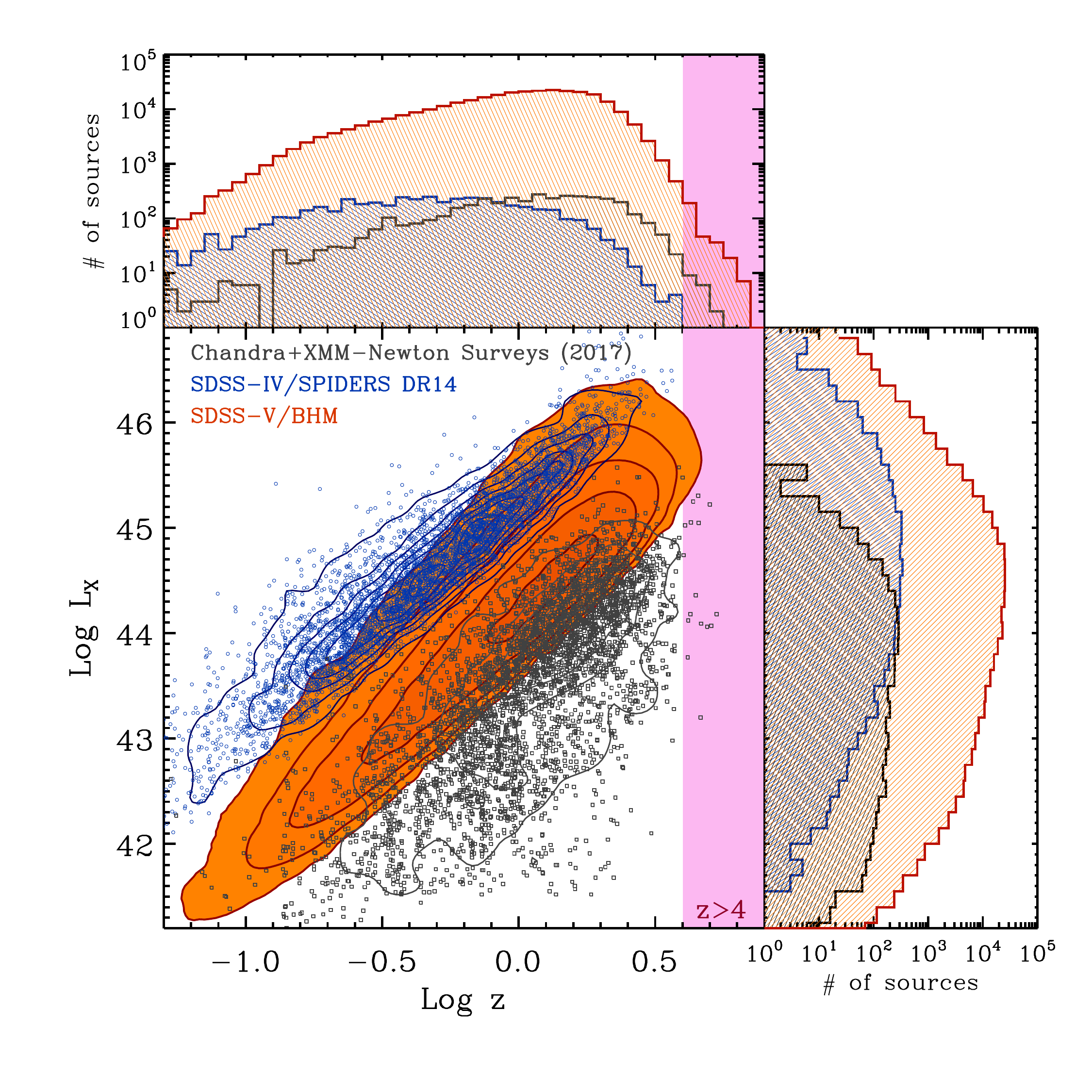}
  \vspace{-0.4in}
  \caption{\footnotesize {\bf Comparison of the X-ray luminosity and redshift coverage of AGN surveys:}  The larger and denser the coverage in this parameter space, the more stringent the constraints that can be placed on the history of accretion onto SMBHs in the Universe. The central plot compares SDSS-V's {\it eROSITA} follow-up program (orange contours; $\sim$316,000 targets) and current state-of-the art AGN surveys: SDSS-IV/SPIDERS (blue; $\sim$4600 targets) and a compilation of deep {\it Chandra} and {\it XMM-Newton} fields (black; CDFS, CDFN, COSMOS, Lockman Hole, XMM-XXL; $\sim$4000 targets). The top and right histograms show the number of AGN expected in each of these samples.
Note the logarithmic y-axes in both histograms: the {\it eROSITA} X-ray and SDSS-V-BHM sample will be about 100$\times$ larger than any
existing sample, spanning a wide range of redshift and luminosities.
The pink shading at $z>4$ highlights the $\sim$10$\times$ improvement in X-ray selected AGN sample sizes at high redshift.}
\label{fig:SPIDERS-Lz} 
\end{figure}

In addition, despite the high X-ray luminosity of nearly all AGN, we do not fully understand the physical origin of the tight coupling between the hot X-ray corona and the ``cold'' accretion disk. This is mostly due to the limited size of X-ray AGN samples compiled with the X-ray telescopes having the necessary sensitivity ({\it Chandra}, {\it XMM-Newton}) but relatively small FOVs.  However, the imminent {\it eROSITA} instrument \citep[extended ROentgen Survey with an Imaging Telescope Array;][]{Predehl_etal_2014} aboard the Spektr-RG mission, set to launch in 2018 with both high sensitivity and a large FOV, will discover as many new X-ray sources in its first twelve months as are known today, after more than 50 years of X-ray astronomy. SDSS-V will provide optical spectroscopic measurements (to about $i_{AB}<21.5$), including identifications and redshifts, of $\sim$400,000 {\it eROSITA} X-ray sources detected in the first 1.5 years of the all sky survey  \citep[i.e., those to a 0.5--2 keV flux limit of $\sim$$2.5 \times 10^{-14}$ erg~s$^{-1}$~cm$^{-2}$; see  Figure~\ref{fig:SPIDERS-Lz} and][]{Merloni_etal_2012}.  This sample will comprise mainly quasars/AGN at high Galactic latitude, but it will also contain X-ray emitting galaxy clusters and X-ray-bright stars (such as compact binaries and flaring late-type stars) in the MW and nearby galaxies. In addition, SDSS-V's BHM will characterize numerous serendipitous discoveries, extreme and rare objects, transients, and other peculiar variables found in the {\it eROSITA} survey \citep{Merloni_etal_2012}, and expand an optical$+$X-ray quasar sample with implications for observational cosmological constraints \citep[e.g.,][]{Risaliti_2015_quasarcosmology}. This combination of X-ray discovery and optical characterization will provide a great leap forward in our description of the X-ray sky, and it will reveal the connections between large, statistical populations of X-ray sources and the cosmic structures in which they are embedded.

\subsection{Local Volume Mapper}
\label{sec:lvm}

We have come to understand that galaxy formation must be a self-regulated process, with energy exchange between stars and interstellar gas occurring at numerous points, both spatially and temporally.  For instance, we know that the rate of star formation (SF) scales with the density of the gas on kiloparsec scales, but feedback and the enrichment of the ISM occurs on the scale of individual stars.  However, many of the observed stellar/gaseous correlations remain basically empirical --- their existence is well-established, but their physical origins are unclear.

\begin{figure}[tbh!]
\centering
\includegraphics[width=\textwidth]{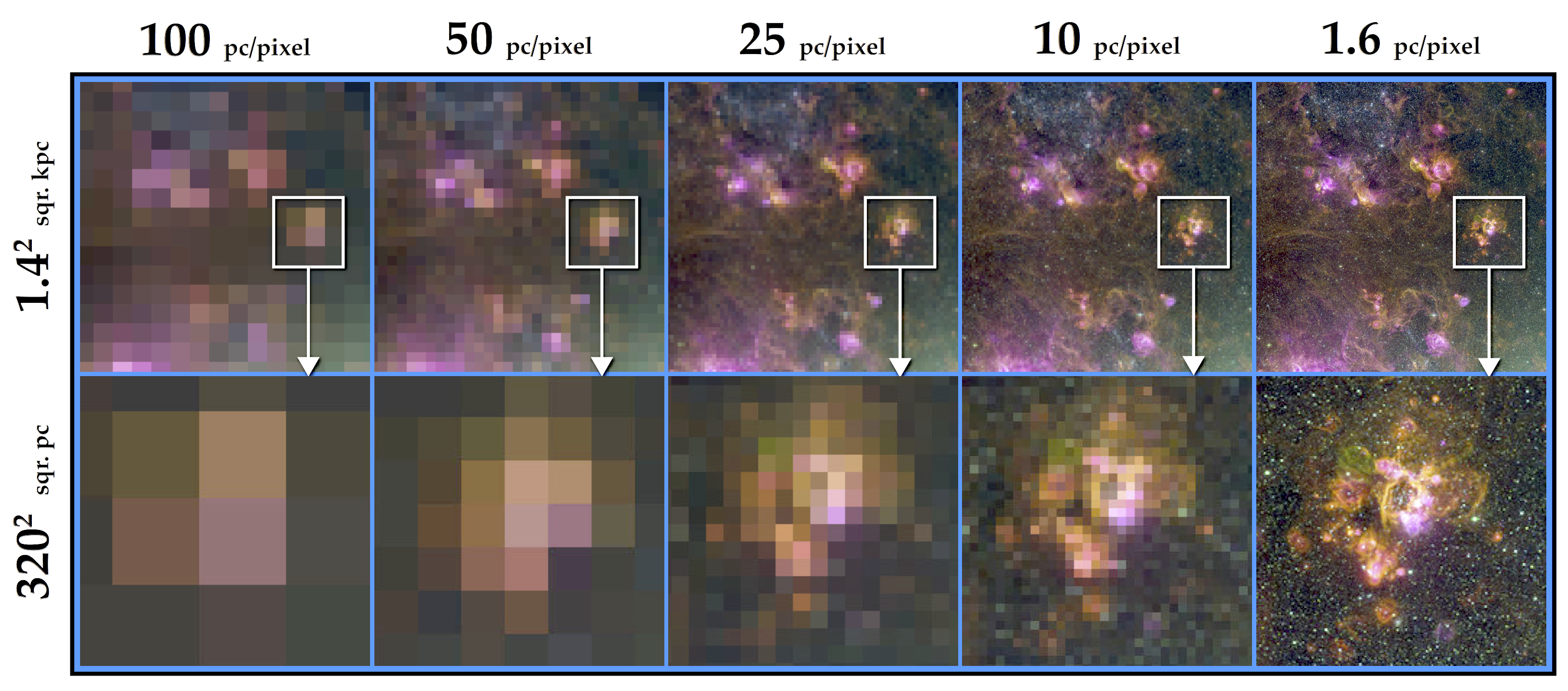}
\caption{
\footnotesize {\bf The dynamic range of spatial scales sampled by LVM:}  
Optical (V-band) and narrow band ([O III], H$\alpha$, [S II]) imaging of the LMC  \citep{Smith+2000}. The sequence starts on the left at the 100~pc resolution typically achieved by the best IFU data available today for external galaxies, and ends on the right with the resolution LVM will achieve in the Milky Way. Note the qualitative change around 25~pc, when networks of shocks and ionization fronts start to appear and then become resolved at $<$10~pc.
\label{fig:LVM_zoom}
}
\end{figure}

SDSS-V's Local Volume Mapper (LVM) will take on this problem by making global ISM maps of Local Group galaxies with a resolution down to the physical scales from which the global correlations arise.  Figure~\ref{fig:LVM_zoom} visualizes how the visible structure in the ISM qualitatively changes at a resolution of 25~pc, below which 50-100~pc sized ``clouds'' can be separated into individual resolved SF knots and the filamentary structures and shock networks between them. This resolution allows diffuse gas to be cleanly separated from ionization fronts and HII regions. IFU studies of MW regions, like Orion, have resolved these structures, but only across arcminute-scale areas corresponding to a few parsecs \citep{Sanchez+2007,Weilbacher+2015,McLeod+2015}. Connecting studies across the pc (sub-GMC) and kpc (galaxy-wide) scales is fundamental to understanding the physics governing star formation, the structure and energetics of the ISM, the baryon cycle, and ultimately, the evolution of galaxies. SDSS-V's LVM will provide this high resolution spectral mapping over large regions of multiple galactic disks, sampling the ISM across a wide range of local galactic environment.

\begin{figure}[t!]
\centering
\includegraphics[width=1.0\textwidth]{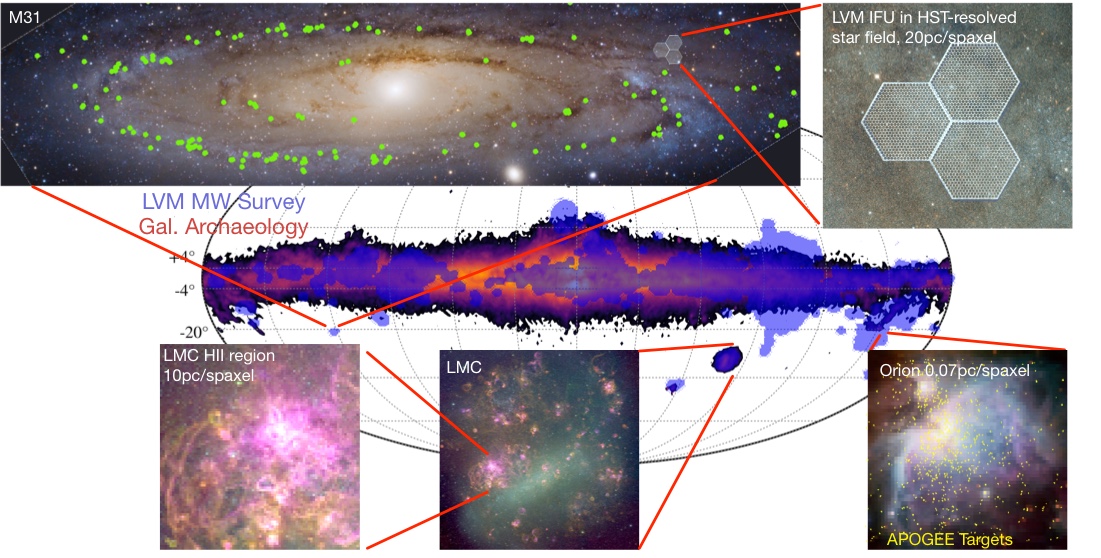}
\caption{\footnotesize
{\bf The LVM footprint and sampling:} In the center, the LVM survey footprint on the MW (blue) is shown on top of the SDSS-V MWM target density map (Section~\ref{sec:mwm}). Zooming into the Orion region (lower right), we see an image of ionized emission sampled at the LVM spaxel size ($<$0.1~pc) and individual stars with high-resolution SDSS-III, -IV, and/or -V spectroscopy with APOGEE in yellow. For the LMC (left/center bottom), we see a continuum $+$ ionized emission image of the 30 Doradus star forming region, sampled at the 10~pc spaxel size LVM will deliver on the Magellanic Clouds. The top panels show the LVM IFU field-of-view over a continuum image of M31. Statistical samples of \HII\ regions (green) observed at 20~pc resolution across M31, and at $\sim 50$~pc resolution across other nearby galaxies, connect small scale physics and large scale galaxy evolution.}
\label{fig:lvm-overview}
\end{figure}

The LVM will provide optical IFU data able to resolve SF structures, giant molecular clouds, HII regions, and young stellar clusters.  These data will span the bulk of the MW disk at 0.1--1~pc resolution, the whole LMC and SMC at 10~pc resolution, M31 and M33 at 20~pc resolution, and Local Volume galaxies at $\sim$20--100~pc resolution,\footnote{We note that the exact sample of more distant objects is not yet finalized and will depend on further tuning of the science case and survey strategy and priorities.} out to a distance of $\sim5$~Mpc (Figure~\ref{fig:lvm-overview}).  This coverage will equal about one full steradian ($\sim$3,300~deg$^2$) of the sky; for comparison, the SDSS-IV MaNGA survey \citep{Bundy+2015} spans about 0.5$^\circ$~deg$^2$ across 10$^4$ low redshift galaxies, and {\it all} of the single-fiber SDSS spectroscopy to date sums to only a few deg$^2$ of sky. LVM will provide as many spectra on the LMC as are contained in all of MaNGA's objects taken together ($\sim$10$^6$). The LVM spectrographs will span 3600--10000\AA\ at a resolving power of $R\sim4000$. This sprawling sky coverage will overlap with datasets providing complementary information at matched spatial resolution. Stellar spectroscopy with accurate typing and abundances from previous APOGEE observations and from SDSS-V itself (Figure~\ref{fig:lvm-overview} and Section~\ref{sec:mwm}) as well as resolved stellar photometry and CMDs (in the Magellanic Clouds, M31 \& M33) will allow us to connect the structures in the ISM to the radiation field and to {\em individual} sources of feedback. Far-IR, sub-mm, and radio surveys probing the dust and H$_2$/\HI\ phases of the ISM connect our observations to molecular clouds and cold gas. X-ray catalogs (e.g., from {\it eROSITA}; Section~\ref{sec:bhm}) indicate the location of X-ray binaries and other additional sources of interstellar ionization.

LVM's extremely wide-field optical spectral mapping adds critical information about the ionized ISM and integrated stellar populations to these complementary datasets, enabling a synthesis of the local $\leftrightarrow$ global physics in galaxy disks.  The key science themes that will be addressed include the connection between ionized gas, star formation, and feedback on multiple physical scales; the extraction of maximum information from the union of resolved and integrated stellar population data; and the geometric structure of the ionized and dusty ISM to better understand chemical abundances and enrichment.  

Within these themes, there are numerous outstanding science questions that can only be answered by wide-area, wide-wavelength surveys.  These range from understanding the energetics, sinks, and sources of stellar feedback; the dependence of star formation on the local galactic environment (e.g., gas density and local shear); the life cycle of GMCs and star-forming complexes; the co-evolution of stellar populations and the surrounding ISM; and the distribution of interstellar metals at high spatial resolution, including the mixing of metals produced in supernovae and AGB stars.  LVM's uniquely well-sampled 2D maps of the gas-phase properties of nearby galaxies (including our own MW) will provide empirical constraints on the ISM at the ``energy injection scale''---the physical scale where stars return energy to their surroundings.  These measurements are critical for constraining theoretical models of star formation and feedback, as well as radiative transfer modeling of ISM structures.

\section{The Implementation of SDSS-V}
\label{sect:SurveyImplementation}

\subsection{Dual-Hemisphere Infrastructure and Instrumentation}
\label{sec:infrastruct_instrument}

SDSS-V will operate multiple telescopes and spectrographs in two hemispheres as a single unified survey across the entire sky. At Apache Point Observatory (APO) in New Mexico sits the 2.5m Sloan Foundation Telescope \citep{Gunn_2006_sloantelescope}, the workhorse of past SDSS generations that will remain dedicated to SDSS-V. At Las Campanas Observatory (LCO) in Chile, the Carnegie Observatories' 2.5m du~Pont telescope \citep{Bowen_1973_duPontTelescope} will be dedicated to SDSS-V for $\gtrsim$300 nights/year. To enable the ultra-wide field IFS mapping, SDSS-V will also use 
a suite of smaller telescopes with apertures ranging from 1~m to 16~cm, to be set up at APO and LCO. 

Both APO and LCO will each house a set of well-established survey instruments for SDSS-V: a near-IR APOGEE spectrograph \citep[300 fibers, R$\sim$22,000, $\lambda=1.5-1.7\mu$m;][]{Wilson_2012_apogee}; a large optical spectrograph \citep[500 fibers, R$\sim$2,000, $\lambda=360-1000$nm;][]{Smee_2013_SdssBossSpectrographs}; and three medium-resolution optical spectrographs feeding a large IFU (2000 fibers, R$\sim$4,000, $\lambda=360-1000$nm). See Figure~\ref{fig:SDSSV_schematic} for a schematic layout of these instruments. For the multi-object spectroscopy used by the MWM and BHM, each 2.5m telescope will be able to observe up to 500 targets at a time, drawing on both the optical and IR multi-fiber spectrographs simultaneously. SDSS-V will rely heavily on the proven SDSS observatory infrastructure, instrumentation, and operations model, including the use of the near-IR APOGEE and optical BOSS spectrographs for the MWM and BHM.  Among the infrastructure innovations, however, are a new robotic fiber positioning (RFP) system for these MOSs and an ultra-wide field IFS.

To meet the requirements for SDSS-V's rapid exposure sequences and high target densities, we will replace the current SDSS plug-plate fiber system on both telescopes with robotic positioners with 500 arms, 300 of which will hold both an optical and an IR fiber (the remaining 200 will hold just an optical fiber).  This system reduces the target reconfiguration time from 20 minutes (by changing plug-plates, with the current system) to under 2 minutes. The fiber positioners can account for atmospheric refraction more easily than drilled plates can, greatly increasing the available observing window for each field and boosting the survey efficiency. Target lists can also be modified on short timescales to allow observations of transients and other targets of opportunity.

The LVM will use a new system that enables IFS over unprecedentedly large areas.  It will expand upon SDSS-IV's existing MaNGA technology, producing an IFS unit based around lenslet-array coupled, tightly packed, abuttable bundles of fibers ($\sim$2000 at each site). These bundles will feed a cluster of three multi-channel spectrographs at each observatory. The LVM instrument could be fed from the Sloan 2.5m, du Pont 2.5m, or the small 1m--16cm telescopes. This flexibility leads to a cascade of possible field sizes, between 1.6$'$ and 25$'$ in diameter, with each fiber subtending between 2.7$''$ and 37$''$ on the sky. Trade studies are ongoing to optimize the exact cascade of apertures and spaxel sizes for all different LVM targets. All of these hardware developments use proven technology for rapid implementation by 2020.

\subsection{Dual-Hemisphere Observations}

The implementation of SDSS-V observations can be divided into two regimes: MOS on the 2.5-meter telescopes (MWM and BHM) and IFS, mostly on smaller telescopes (LVM).  There will be small differences in the MOS observations between LCO and APO, such as plate scale \citep[e.g.,][]{Zasowski_2017_apogee2targeting}, but significant effort will be put towards making data from the two hemispheres as homogeneous as possible for the end user.  Observations in both hemispheres, for the three Mappers, will run almost entirely in parallel for the duration of the whole survey.

The survey planning for MOS observations is currently based on an exposure quantum of $t_{\rm exp} \approx 15$~min (in contrast to the $\approx$1~hour exposures of SDSS-I through -IV).  The details of  $t_{\rm exp}$ are still undergoing study and optimization, but the requirement of rapid, short exposure is clear: it enables significantly higher survey efficiency over a wide range of target brightnesses.  The RFP system will support the rapid and flexible fiber (re-)allocations necessary for this increased efficiency (see Figure~\ref{fig:SDSSV_programs} for the projected density of SDSS-V targets).  Any given set of 500 MOS targets can be scheduled for a combination of optical spectroscopy (up to 500 fibers) and/or IR spectroscopy (up to 300 fibers).  Repeat ``visits'' to the same field, where each ``visit'' is $t_{\rm exp}$ long, will be used to a) increase the final density of targets by observing new sources, b) boost the S/N of source spectra by adding exposure time, regardless of the observing epoch, and c) obtain time-resolved spectroscopy, where the temporal cadence matters (e.g., binary stars and quasar variability).  The joint availability of these options, enabled by the RFP and the fact that both MOS instruments will share the focal plane at all times, opens up many paths to survey optimization.

As described in Section~\ref{sec:infrastruct_instrument}, the IFS observations of LVM will span a range of field sizes and plate scales (up to $37^{\prime\prime}$ per spaxel), allowing flexible optimization of sensitivity versus spatial resolution.  Since these observations depend largely on the smaller SDSS-V telescopes, and use a separate cluster of spectrographs, LVM will generally operate concurrently with BHM and MWM in both hemispheres.  The large-area components, such as the mapping of the MW disk and Magellanic Clouds, will be on the smallest telescopes and will be as automated as possible.  The calibration scheme for the MW survey will be different from other IFS surveys, including the other LVM components, because the large spatial extent of the targets covers the whole telescope FOV and prevents simultaneous observations of sky and spectrophotometric standards. A novel calibration strategy using simultaneous observations of calibration fields with a dedicated calibration telescope will be adopted, the details of which are still being finalized.

\subsection{What sets SDSS-V Apart? The 2020 Landscape of Spectroscopic Surveys}

SDSS-V's potential for multiplying the science to come from the {\it Kepler}, {\it Gaia}, {\it TESS}, and {\it eROSITA} datasets is apparent from these missions' summaries in Table~\ref{tab:space_missions}. {\it Kepler} and {\it TESS}'s magnitude ranges are very well matched to SDSS-V's IR/APOGEE spectroscopy, with MWM providing precise stellar parameters, abundances, and multi-epoch velocities (to $\sim$30~m~s$^{-1}$) for their targets. 
SDSS-V's optical/BOSS spectroscopy is well matched for the magnitude range of unextinguished {\it Gaia} sources, providing basic spectroscopic characterization at the faint end and excellent characterization at the bright end. The high optical extinction for most sources in the GGS program at low Galactic latitudes means that most of the spectroscopic GGS targets ($H<11$) span the same optical magnitude range ($14<G<19$) as most of {\it Gaia}'s sources with sensibly precise astrometry. 
While a good fraction of {\it eROSITA}'s detections made later in its multi-epoch X-ray survey will be followed up from 4MOST, SDSS-V will perform the first uniform optical MOS follow-up of hundreds of thousands of X-ray sources discovered in the first 1.5 years of {\it eROSITA}.

\begin{table}[bh!]
\scalebox{0.9}{
\begin{tabular}{ |p{2cm}||p{7.9cm}|p{2cm}|p{4.4cm}|  }
 \hline
 \multicolumn{4}{|c|}{{\bf Wide-field Space Missions Enhanced by SDSS-V}} \\
 \hline
 Mission & Science Goals / Data Products  & Timeframe & Primary Mag Range\\
 \hline
 \hline
\multirow{2}{*}{{\it Kepler/K2}}   & (transiting) exoplanets \& stellar astrophysics
\newline seismology from precision lightcurves & \multirow{2}{*}{2009--2018} & \multirow{2}{*}{$m_{V}\sim $ 7--17; selected fields} \\
 \hline
  \multirow{2}{*}{{\it Gaia}}  & positions, distances, motions from astrometry; \newline basic stellar parameters  & \multirow{2}{*}{2013--2020} & \multirow{2}{*}{$m_{G}\sim $7--19; all-sky} \\
  \hline
  \multirow{2}{*}{{\it TESS}}   & (transiting) exoplanets \& stellar astrophysics 
  \newline  seismology from precision lightcurves & \multirow{2}{*}{2018--2022} & \multirow{2}{*}{$m_{i}\sim $ 8--14; $\sim$all-sky} \\
 \hline 
 \multirow{2}{*}{{\it eROSITA}} & \multirow{2}{*}{X-ray fluxes  \& spectra} & \multirow{2}{*}{2018--2022} &  $f_{\rm 0.5-2keV} >10^{-14}$ \newline erg~s$^{-1}$~cm$^{-2}$; $\sim$all-sky \\
 \hline
\end{tabular}
}
\caption{\footnotesize\linespread{1.2}{Surveys by (mostly imaging) space missions in 2020-2025 timeframe}}
\label{tab:space_missions}
\end{table}

SDSS-V is different from the array of current or imminent ground-based MOS surveys (Table~\ref{tab:spectroscopic_surveys}), enhancing them by opening up a unique part of ``survey'' parameter space.
This bold claim requires quantification and discussion. First, we note that SDSS-V is not the optimal survey facility for science with sources that are faint, are not very red (intrinsically or because of low extinction), and do not call for multi-epoch or all-sky observations. 
Such science includes much of cosmology, large-scale structure, high-redshift galaxy evolution, and detailed studies of substructure in the Galactic halo. This science is at the heart of many of the surveys in 
Table~\ref{tab:spectroscopic_surveys}, whose outstanding design makes them better able to serve these science goals than SDSS-V can.

But when it comes to the scientific heart of SDSS-V's multi-object spectroscopy --- all-sky, multi-epoch, dust-penetrating near-IR spectroscopy of many millions of targets --- Table~\ref{tab:spectroscopic_surveys} shows that SDSS-V has a clearly unique place: it is the only survey to come close to full 4$\pi$ sky coverage; only it will cover the entire inner Galaxy at low latitudes, where most Galactic stars live; 
and only it will have comprehensive multi-epoch observations across the sky.  For many surveys, the science goals 
argue against ``covering the sky'' with short exposures. 

While {\it Gaia} has a spectroscopic survey component (with its RVS instrument), these data have relatively small spectral coverage (only 10\% the spectral resolution elements of each of SDSS-V's spectrographs), and its flux limit for high-quality spectra will be a factor of 100 brighter than for SDSS-V's optical stellar spectra. The WEAVE survey, starting a couple of years before SDSS-V, will come closest to SDSS-V's ambition and will form a crucial benchmark for SDSS-V to build upon by, e.g., providing a well-established ``ground truth'' in small portions of the Galactic plane. Starting toward the anticipated end of SDSS-V in 2024, 4MOST will provide important complementary data in the form of deeper optical spectra, including a focus on the inner Galactic disk and bulge. These facility comparisons show that SDSS-V will neither compete with nor follow these surveys: it will focus on exploring a new and different regime.

\begin{table}[h]
\scalebox{0.74}{
\begin{tabular}{|p{2.8cm}||p{1.7cm}|p{1.6cm}|p{1.2cm}|p{2.3cm}|p{2.9cm}|p{1.75cm}|p{2cm}|p{1.8cm}|}
 \hline
 \multicolumn{9}{|c|}{{\bf Spectroscopic Survey Facilities by 2020--2025}} \\
 \hline
 Survey (facility)&$N_{\rm target}$ & $R_{\rm spectra}$ & $N_{\rm multi}$ & $\lambda[\mu m]$ & $\Omega_{\rm sky}$ & $N_{\rm epoch}$ & Timeframe & $m_{\rm primary}$\\
 \hline
 \hline
 \multirow{2}{*}{{\bf SDSS-V}} & \multirow{2}{*}{$7\times 10^6$} & {\bf 22,000\newline 2,000} & \multirow{2}{*}{{\bf 500}} & {\bf 1.51--1.7 \newline 0.37--1.0} & \multirow{2}{*}{{\bf $4\pi$}} & \multirow{2}{*}{{\bf 1--174}} & \multirow{2}{*}{{\bf 2020--2024}} & {\bf $m_H \lesssim 13.4$ \newline $m_i \lesssim 20$} \\
 \hline
 \hline
Gaia (RVS) & $8\times 10^6$ & 11,000 & --- & 0.85--0.87 & $4\pi$ & $\sim$60 & 2013--2020 &$m_G \lesssim 12$ \\
 \hline
Gaia-ESO & $0.1\times 10^6$ & $17,000$ & 140 & 0.55 \& 0.85 & $0.02\pi$ & $\sim$1 & 2013--2018 & $m_G\lesssim 17$ \\
 \hline
GALAH & $0.8\times 10^6$ & $28,000$ & 400 & 0.40--0.85 & $\pi$, $|b|\ge 10$ & $\sim$1 & 2015--2020 & $m_G \lesssim 13$\\
 \hline
\multirow{2}{*}{WEAVE} & \multirow{2}{*}{$0.8\times 10^6$} & $5,000 \newline 20,000$ & \multirow{2}{*}{1000} & \multirow{2}{*}{0.37--0.9} & \multirow{2}{*}{$\sim$$\pi$} & \multirow{2}{*}{$\sim$1--2} & \multirow{2}{*}{2018--2023} & \multirow{2}{*}{$m_G \lesssim 19$} \\
 \hline
DESI & $4\times 10^7$ & 3,000 & 5000 & 0.36--0.98 & 1.35$\pi$, $|b|\ge 25^\circ$ & 1--4 & 2019--2024 & $m_r \lesssim 23$ \\
 \hline
LAMOST & $8\times 10^6$ & 1,800 & 4000 & 0.4--0.9 & $0.5\pi$ & $\sim$1 & 2010--2020 & $m_G \lesssim 16$ \\
 \hline
\multirow{2}{*}{4MOST} & \multirow{2}{*}{$10\times 10^6$} & 5,000 \newline 20,000 & 1600 \newline 800 & \multirow{2}{*}{0.4--0.9} & \multirow{2}{*}{$1.5\pi$} & \multirow{2}{*}{1--2}& \multirow{2}{*}{2023--2028} & $m_r \lesssim 22$ \newline $m_V \lesssim 16$ \\
\hline
APOGEE-1\& 2 & $5\times 10^5$ & $22,000$ & 300 & 1.51--1.7 & $0.5\pi$ & $\sim$1--30 & 2011--2019 & $m_H \lesssim 12.2$ \\
\hline
PFS & $1\times 10^6$ & 3,000 & 2400 & 0.4--1.6 & $0.05\pi$ & $1$ & 2018--2021 & $m_i \lesssim 23$ \\
\hline
\multirow{2}{*}{MOONS} & \multirow{2}{*}{$2\times 10^6$} & $5,000 \newline 20,000$ & \multirow{2}{*}{1000} & \multirow{2}{*}{0.6--1.8} & \multirow{2}{*}{$0.05\pi$} & \multirow{2}{*}{1} & \multirow{2}{*}{2020--2025} & $m_g \lesssim 22$\newline $m_H \lesssim 17$\\
\hline
 
 \hline
\end{tabular}
}
\caption{\footnotesize\linespread{1.2}{Surveys by spectroscopic facilities in 2020--2025 timeframe.  For each survey, the columns show anticipated number of unique targets ($N_{\rm target}$), spectral resolution ($R_{\rm spectra}$), multiplex number of the MOS ($N_{\rm multi}$), wavelength range ($\lambda[\mu m]$), sky coverage ($\Omega_{\rm sky}$), timescale for survey, and approximate target magnitude range ($m_{\rm primary}$). We note that these parameters, especially $m_{\rm primary}$, for many of the surveys are approximate, based on current public documentation and comparable survey subprograms.}}
\label{tab:spectroscopic_surveys}
\end{table}

\subsection{Data, Collaboration and Science: the SDSS Legacy}

SDSS-V aims to be the successor to the highly successful SDSS I--IV.\footnote{\href{https://www.sdss.org}{www.sdss.org}} The main science programs that we have outlined here reflect the result of a nearly two-year process of open discussion to identify science directions where 2m-class (or smaller) spectroscopic telescopes can produce transformational astrophysics. The science foci that this process identified offer enormous promise, even as they signal a re-weighting of the cosmology and large-scale structure programs that have heavily influenced the SDSS framework for the better part of 20 years.  This evolution makes sense:  the frontier in those particular fields has moved to larger telescopes and other wavelengths, while stellar and galactic astrophysics have been rejuvenated, following, e.g., the {\it Kepler} and (anticipated) {\it TESS} space photometry revolution. 

While in its astrophysical core goals the transition to SDSS-V constitutes a larger change than the gradual science evolution of SDSS I$\rightarrow$IV, SDSS-V will continue in the scientific and collaborative spirit that has made SDSS one of the leading international astronomical collaborations.  These basic operational tenets of our consortium of foundations, institutions, and individuals include: fully open collaboration within the consortium, with no ``reserved'' data or topics for individuals or sub-groups; inclusive co-authorship policies; high priority given to creating well-calibrated and well-documented data sets, along with their regular release to the global science community; an inclusive, collaborative climate that allows junior researchers to carry out high-impact SDSS science and build their careers; a dedication to student research, education, and public outreach; and an organically evolving consortium, with continual opportunities to evolve the science and survey strategy.

SDSS's success and influence in the astronomical community 
reflect both the high quality and broad diversity of its data and the exemplary spirit and functioning of its collaboration. SDSS-V is committed to continuing these traditions while making strides to further improve upon the supportive environment and the diversity and inclusiveness of leadership roles within its organizational structure.  

\section{The Path to SDSS-V: Project Development and Collaboration Building}
\label{sect:DevelopmentCollaboration}

In the tradition of its predecessors, SDSS-V is an ambitious survey with an ambitious timeline for construction and implementation.  Our goal is to begin operations at both APO and LCO with our new RFP system and IFS instrumentation (Figure~\ref{fig:SDSSV_schematic}) at the end of SDSS-IV, currently planned for mid-2020.  We are planning a five-year survey, pending fund-raising success.  In practical terms, this requires: 1) construction of two RFP systems to feed the BOSS and APOGEE spectrographs; 2) construction of two large IFS bundles and their marriage to six spectrograph banks; and 3) refining and finalizing our survey strategy and associated pipeline enhancements.  To meet this demanding schedule, we focus on known technological solutions (rather than designing new systems) for much of the hardware described above.  

Equally critical to the success of SDSS-V is the building of the SDSS-V scientific collaboration --- a world-wide network of institutions working together towards the goals described here. The extraordinary science outlined above {\it simply does not happen without the participation and support of institutional partners}.  There are over 50 current partners in SDSS-IV, ranging from associate institutions (with a restricted number of individual participant slots) to full institutional members.  As of the writing of this document, SDSS-V has early participation from 19 institutional partners, including full membership from The Carnegie Observatories, MPIA, MPE, and the University of Utah.  Full membership affords every member of an institution full data rights to the survey data and grants the institution a vote on the Advisory Council, the high-level body responsible for decision-making in the survey.  The cost of full membership in SDSS-V is \$1.15M~USD.  SDSS-V members (at all levels of membership) enjoy proprietary access to all survey data for a period of two years.  In addition to this access, members are crucial in the survey planning and have access to the planet-wide collaboration network.

\section{SDSS-V: A Pathfinder for Panoptic Spectroscopy}
\label{sec:future} 
Several aspects of SDSS-V position it to pioneer high-quality panoptic spectroscopy: industrial-scale, all-sky, multi-epoch spectroscopy, with $\Delta t_{\rm epoch}$ ranging from 20 minutes to 20 years; contiguous wide-field IR spectroscopy to map the obscured Galaxy; optical IFS that covers an appreciable fraction of the sky, etc.  SDSS-V is also an important step on the path towards a spectroscopic counterpart to LSST.  Such a facility has been suggested and requested in numerous reports \citep[e.g.,][]{Najita_2016_maximizeLSST,Ellis_2017_spectroLSST}.  This future system would have the light-gathering power of 8m wide-field telescopes to reach many magnitudes deeper, thousands of fibers to survey the fainter sources revealed by large telescopes, and a surveying speed and operations model that allows systematic, all-sky time-domain spectroscopy not only of variable sources, but also of transient ones. While SDSS-V is clearly focused on bright sources, we will work --- both in the preparation and the implementation of SDSS-V --- towards enabling target-of-opportunity spectroscopy of transient phenomena. The lessons learned from this pioneering survey will prove invaluable in planning for the scientific merits and requirements for a spectroscopic counterpart to LSST.

\bibliographystyle{yahapj}
\bibliography{arxiv}{}

\end{document}